# Complementary Textures.
# A Novel Approach to Object Alignment in Mixed Reality*

Alejandro Martin-Gomez, Alexander Winkler, Rafael de la Tijera Obert,
Javad Fotouhi, Daniel Roth, Ulrich Eck, and Nassir Navab

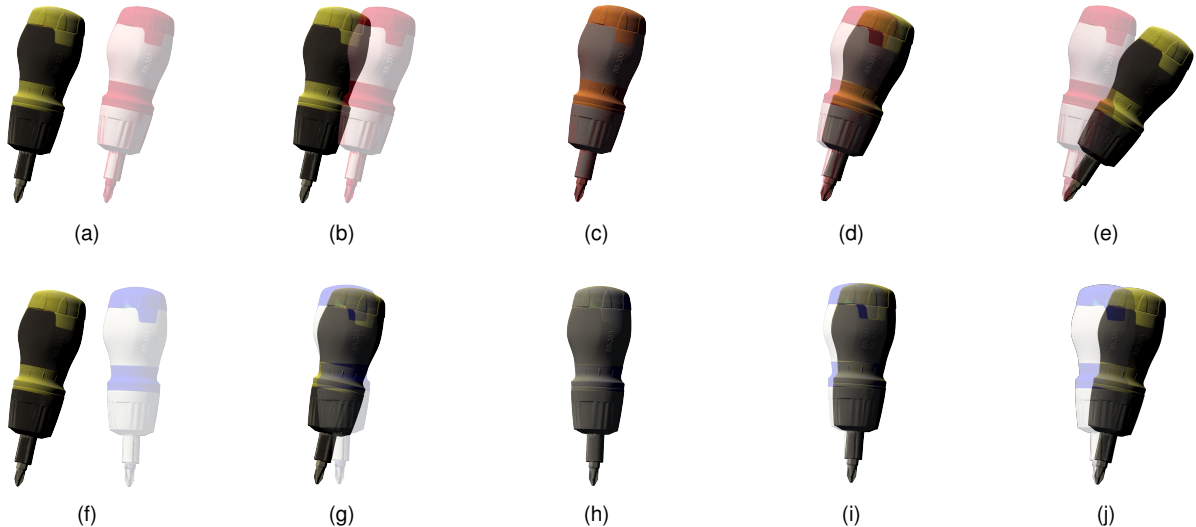

Fig. 1. COMPLEMENTARY TEXTURES for object alignment in Mixed Reality. This concept utilizes the textural surface of a real object to generate a virtual replica with a complementary pattern to assist users during alignment tasks. The replica is designed to generate highly salient error visualization when the objects are not aligned in position (a),(b),(f),(g), or orientation (d),(e),(i),(j). In addition, it allows to visualize a new texture (c) or a homogeneous single-colored object (h), when proper alignment is achieved.

**Abstract**—Alignment between real and virtual objects is a challenging task required for the deployment of Mixed Reality (MR) into manufacturing, medical, and construction applications. To face this challenge, a series of methods have been proposed. While many approaches use dynamic augmentations such as animations, arrows, or text to assist users, they require tracking the position of the real objects. In contrast, when tracking of the real objects is not available or desired, alternative approaches use virtual replicas of real objects to allow for interactive, perceptual virtual-to-real and/or real-to-virtual alignment. In these cases, the accuracy achieved strongly depends on the quality of the perceptual information provided to the user. This paper proposes a novel set of perceptual alignment concepts that go beyond the use of traditional visualization of the virtual replicas, introducing the concept of COMPLEMENTARY TEXTURES to improve interactive alignment in MR applications. To showcase the advantages of using COMPLEMENTARY TEXTURES, we describe three different implementations that provide highly salient visual cues when misalignment is observed; or present semantic augmentations that, when combined with a real object, provide contextual information that can be used during the alignment process. The authors aim at opening new paths for the community to explore rather than describing end-to-end solutions. The objective is to show multitude of opportunities such concepts could provide for further research and development.

**Index Terms**—Mixed Reality, Augmented Reality, Object Alignment, Visualization, Perception

✦

## 1 INTRODUCTION

One of the multiple applications for which MR has shown to be highly valuable is to provide users with visual guidance during the manipulation and placement of real objects. This topic has been widely explored in multiple industrial scenarios that involve assembly and manufacturing [13, 21, 22, 28, 29, 31], training [11, 20, 30], or maintenance [5]; as well as in medicine to assist physicians during the insertion of surgical instruments [1, 10], or to support users with the set up and positioning of medical devices [6, 9, 15–17] and surgical robots [7].

To assist users of MR applications during manipulation and placement tasks, a series of methods have been proposed. Some methods guide the user by presenting visual cues in the form of virtual animations, arrows, or text that are updated dynamically during the performance of the task. These methods require continuous tracking of the real objects to compute the spatial transformation between their current and desired pose. When tracking the real objects cannot be fulfilled, or it is not desired, alternative methods provide static augmentations in the form of virtual replicas of their real counterparts that are used to indicate the desired pose of the objects to align. These methods

• *Alejandro Martin-Gomez (E-mail: alejandro.martin@jhu.edu), Rafael de la Tijera Obert, and Nassir Navab are with the Laboratory for Computational Sensing and Robotics, Johns Hopkins University.*
• *Alexander Winkler, Ulrich Eck, and Nassir Navab are with the Chair for Computer Aided Medical Procedures, Technical University of Munich.*
• *Javad Fotouhi is with Philips Research North America.*
• *Daniel Roth is with the Human-Centered Computing and Extended Reality Lab, Friedrich-Alexander-Universität Erlangen-Nürnberg.*

ease the manipulation of the real objects as there is no need to attach optical markers or to integrate external cameras to track them. However, the alignment accuracy that can be achieved with these methods strongly depends on the quality of the information provided by the visual augmentations.

Traditional visualization techniques used in such methods present virtual replicas of the objects of interest in the form of solid [13, 21, 28], semitransparent [3, 20, 22, 30, 31], wireframe [5, 29], or point cloud [6, 9] representations. However, existing studies suggest that the choice of visualization technique influences the user's performance in such alignment tasks [18]. Moreover, while these visual representations can be used to infer alignment errors during task performance, they are not designed to highlight potential misalignment but simply to indicate the desired final pose. In addition, accurate estimation of the spatial position of virtual content in MR scenarios has proven to be a challenging perceptual task and is still an open research topic. This is not surprising as humans are not accustomed to visualizing real and virtual objects simultaneously and they are also not used to aligning an object using its virtual replica.

In this paper, we introduce the concept of COMPLEMENTARY TEXTURES as a novel method that exploits the textural surface patterns of real objects to generate virtual replicas that provide rich visual cues to improve object alignment in MR. In contrast to existing approaches, we modify the appearance of the virtual objects (e.g. its color and texture) to present highly salient visual cues, when real and virtual objects are not aligned. This enables us to go beyond the simple generation of contours, wireframes or semitransparent visualizations, and to facilitate visual alignment tasks (see Figure 1). To our knowledge, this is the first attempt towards using the textural properties of the real objects to modify the representation of the replica in order to optimize the perceptual alignment tasks in MR.

To further explore this concept, a more detailed review of the related work is presented in Section 2. Section 3 introduces three different variations of the proposed COMPLEMENTARY TEXTURES that take advantage of photometric, geometric and semantic properties of the real object, respectively. To gain further insights, some proof-of-concept implementations of the COMPLEMENTARY TEXTURES and their counterparts using traditional approaches are provided in Section 4. Lastly, inspiring and promising results of our experiments are discussed in Sections 5 and 6. Please note that this new concept aims at taking advantage of specific visual, geometric or semantic information of the object of interest. Therefore we expect that different classes of objects require different sets of automatic methods, and the handful of solutions presented here, only scratch the surface. The examples presented in this paper may be considered as a first set of possible implementations while in each field of application, designers could propose specific methods for automatic generation of most relevant COMPLEMENTARY TEXTURES.

## 2 RELATED WORK

Object placement and alignment of real and virtual objects is a common task and a requirement for several applications that involve the use of MR. In this regard, interactive approaches can assist the user during the manipulation of the objects by: i) providing explicit instructions in the form of dynamic augmentations such as animations, arrows, or text, or ii) implicitly indicating the desired placement position by using a virtual representation of the object to be aligned.

Providing explicit instructions has proven to be valuable when performing procedural tasks that require object alignment in several fields. In the context of industrial applications, early work from Caudel and Mizell [29] presented an Augmented Reality (AR) system used to provide visual assistance to users during aircraft manufacturing tasks. Feiner et al. [5], introduced a system to aid users during printer repairing. Work by Reiners et al. [21] provided visual guidance during automotive-related assembly tasks. Similar approaches, as the one proposed by Zauner et al. [31] presented an active system able to provide detailed instructions during the performance of an assembly task. Henderson et al. [11] presented a prototype able to provide dynamic augmented instructions using arrows and text during the performance of a procedural task. This work was later extended by Oda et al. [20], introducing a system that enabled local users to receive instructions from an expert during the performance of the assembly task. In addition, Westerfield et al. [30] presented a system providing adaptive instructions to aid users during training for motherboard assembly. This system gave detailed instructions in the case that tasks were not executed adequately. Similar approaches have been used in medical scenarios to provide surgical guidance. Liu et al. [17] introduced a method to provide guidance during hip resurfacing procedures using arrows to indicate a desired drilling trajectory. Liebmann et al. [16] used a tracked tool to indicate the desired trajectory for the placement of pedicle screws in spine procedures.

On the other hand, when the position of the real objects is not available, one can use static visualizations to indicate the desired positions of the objects. A common approach to provide this information consists in presenting a virtual replica of the object to be aligned by the user. These replicas, are normally visualized in the form of solid, semitransparent or wireframe versions of the real object. Tang et al. [28] evaluated the effectiveness of using AR to assist users during assembly tasks with building blocks. This approach presented a solid virtual replica of the object to be assembled in combination with a static arrow indicating the desired position. A similar approach presented by Robertson et al. [22] investigated the effects of placing the virtual information outside of the task area during assembly tasks using semitransparent virtual replicas of building blocks. Along this line, Khuong et al. [13] proposed a system to compare the effectiveness of providing guidance, when the virtual replicas of the models were presented on top and next to the real objects. In the medical context, the work presented by Hajek et al. [9] used marker-less tracking to record a point cloud of a medical imaging device at a desired pose that was later used to assist clinicians during the re-alignment of the device. The use of point clouds to visualize the objects of interest was also used by Fotouhi et al. [6] for the alignment of implants in total hip arthroplasty procedures. In addition, Fotouhi et al. [7] introduced the concept of Reflective-AR Displays that provided multiple reflected views to visualize the changes of virtual content from a first view perspective during the alignment of surgical robotic arms without the need of tracking the real objects.

Achieving accurate alignment, when only visualizing the virtual replica, represents a challenging task and is still an open topic in AR. In this scenario, the accuracy in the alignment strongly depends on the quality of the visual information presented to the user. This type of visualization can lead to incorrect estimation of depth, distances, or shape, and affect user's performance due to the perceptual ambiguities associated with AR visualization [19]. Martin-Gomez et al. [18] presented a comparison of traditional visualization techniques used for alignment in MR. Results from their study suggest that the choice of visualization technique influences the user's performance during alignment tasks. Even though this work presented an experimental setup using a virtual environment, the techniques compared in the study were representative of the most common approaches used to provide implicit visual guidance.

## 3 METHODS

In this paper we introduce the concept of COMPLEMENTARY TEXTURES as a novel alternative to existing methods for providing visual guidance during the performance of alignment tasks. To the best of our knowledge, this is the first time in object alignment to transform the texture of the replica into its complement rather than a copy of the object model. In this section, we explain how the COMPLEMENTARY TEXTURES can assist users during the alignment of virtual and real objects in MR applications. Moreover, we provide a detailed description of this concept and present a set of methodologies that consider its exemplary implementation enabling users to judge whether real and virtual objects are properly aligned or not.

To achieve this, a virtual representation of the object to be aligned is presented using a simple yet effective visualization technique that considers the texture and geometry of the real object. The virtual replica is modified in such a way that it provides strong visual cues when misalignment occurs. This enables the user to perceive alignment errors without the need of tracking the pose of the object that is being manipulated and without providing specific instructions that commonly include numerical and visual information associated to relative position and orientation errors.

### 3.1 Definition

We define COMPLEMENTARY TEXTURES as a novel visualization technique that exploits preexisting textural surface patterns on real objects to improve object alignment in MR environments. This technique provides complementary visual cues to the user during real-to-virtual or virtual-to-real object alignment, and aims at adopting perceptual properties, similar to those associated with Gestalt psychology [14, 23]. This school of psychology establishes that the parts of a geometrical shape, or even the tones of a melody, interact in such a way that they produce a perceived whole that is distinct from the sum of its parts [23]. These components represent qualities of an experience that are not inherent in its components. Even more, the phenomenon of *amodal perception* that derives from *Gestalt psychology*, indicates that it is possible to perceive spatial structure even when there is absence of a physical stimulation as described by Lehar [14]. Such principle has been explored by Breckon [2] in the context of 3D computer vision to explore its application for volume completion.

In this regard, and in the context of object alignment in MR, the combination of textures of each object involved in the task (i.e., the real object and its respective virtual replica), provide useful information that goes beyond the information conveyed by observing each individual object. The complementary information can be delivered to the user, for example, by the generation of **photometric** complements that lead to the observation of a homogeneous surface pattern, when real and virtual objects are aligned. Alternative modalities can use **geometric** elements drawn over the surface of a virtual replica to facilitate the alignment task. In addition, **semantic** augmentations relevant to the objects to align can be perceived as familiar by the users during the alignment task and therefore be used to provide *implicit-guidance*.

In the following, we introduce potential variants and benefits of our proposed replica's texture design and visualization methods for alignment.

#### 3.1.1 Photometric Complements

The **photometric complementary textures** consist of a pair of interrelated color appearances that, once properly aligned, form a homogeneous single-colored object (Figure 1a-1e). These color appearances correspond to the preexisting textural surface pattern of a real object and a virtual texture generated using the same layout of shapes as that of the real object's surface pattern, but inverting its colors.

In this context, a photometric complement of an image consists of any alternative image that, when combined or added with its original counterpart, cancels out (i.e., loses hue) or creates a novel uniformly colored patch. Consequently, the photometric complementary textures generated for three-dimensional objects could result in a single or multicolor form composition when the real and virtual objects are aligned as depicted in Figures (1a)-(1e) and (1f)-(1j), respectively. Due to their complementary properties, the colors of the textural surfaces of the real and virtual objects create the strongest contrast when placed next to each other. In addition, when the alignment error between the objects is small, this type of complementary textures provides visual cues similar to the results obtained when using edge detection algorithms. These visual cues are observable even when the alignment errors occur in depth with respect to the user's viewing direction.

In terms of visualization, the way photometric complementary textures are presented to the observer varies depending on the display technology used to deliver the augmentations and has to be considered when designing the texture. In this regard, while using Optical See-Through (OST) Head-Mounted Displays (HMDs) provides an inherent additive aspect, when the virtual and real textures overlap, the use of Video See-Through (VST) HMDs would result in the virtual object occluding its real counterpart. Therefore, when using VST-HMDs, two approaches can be implemented to generate the photometric complementary textures depending on the visual outcome that is expected once the objects are aligned. On the one hand, if the resulting homogeneous pattern is desired to be the same color as the predominant chromaticity observed in the real object, the complementary texture would inpaint only the pixels in the virtual texture that are different to the predominant color, while the rest of the pixels would be rendered as fully transparent. On the other hand, if the application requires a colorization of the patch generated after alignment, adjusting the transparency values and blending the virtual texture with the background would produce a similar result as the one observed when using OST-HMDs.

#### 3.1.2 Geometric Complements

The texture of the virtual replica can also be presented as a **geometric complementary texture** consisting of geometrical primitives including not only edges but also diagonals, bisectors, arcs, inscribed and circumscribed circles and polygons, as well as more complex patterns such as fractals. These primitives can often be computed automatically from the geometry of the object of interest (see Figure 2).

The geometric complements provide multiple visual cues to improve the perception of alignment and help reducing related errors. One of the advantages of this modality of complementary textures is that it provides visual feedback without leading to a visually occluded environment when the real and virtual objects overlap. Therefore, they can be used in scenarios where the alignment task requires the user to be aware of the real environment as it is in the case of several industrial and medical applications. This capability can be thought to be similar to the properties observed when using edge-based visualizations. However, it is important to mention that the geometric complementary textures emphasize certain geometrical structures that can be used during the alignment task instead of simply rendering the edges of the objects of interest. Thus, the cues provided by the geometric complements can be used to infer errors during task performance and to improve the alignment of the objects of interest.

Note in particular that when using geometric complements, any misalignment between the real and virtual objects can be perceived globally and locally even for textureless objects. Once proper alignment is achieved, the resulting overlap of the real and virtual content in the scene leads to the visualization of a harmonious scene in which the virtual and real content seem to be part of the same object. An exemplary implementation of this concept is depicted in Figure 2.

#### 3.1.3 Semantic Complements

As an alternative to the previous modalities, the virtual replicas used during the alignment task can be presented to the users in the form of **semantic complementary textures** (see Figure 3). This type of textures extends from its geometric and photometric properties to include semantic content by presenting visual cues that are familiar to humans in the context of the objects to align. Thus, the virtual texture is designed to augment the corresponding real object in a way that humans perceive it as a natural addition of the object to align.

Simple examples of this modality could be aligning a human face using virtual objects in the form of sunglasses, hats, or masks. An exemplary image of this concept, is illustrated in Figure 3 where a semantic complement of the Mona Lisa (in the context of the COVID-19 situation) consists of a face mask and a pair of surgical gloves that are used to provide visual cues during an alignment task. This concept can even be extended to the completion of compound letters or words from individual letters or word components as a semantic completion.

### 3.2 Effect

The presented variants of complementary textures have in common that they describe a single appearance, which can be separated into two distinct components. Each component has a meaning of its own, so that the two parts of the complementary textures can be readily understood by an observer when inspected individually. To give an example, a white triangle on a black background and a black triangle on a white background are complements, but for the observer even a single component forms a message. The same principle is true for an image of a landscape and the same image with its colors replaced by its complementary colors. Thus, both images evoke the understanding of the same landscape. Just as well, images of a face and of a pair of sunglasses are both understandable individually.

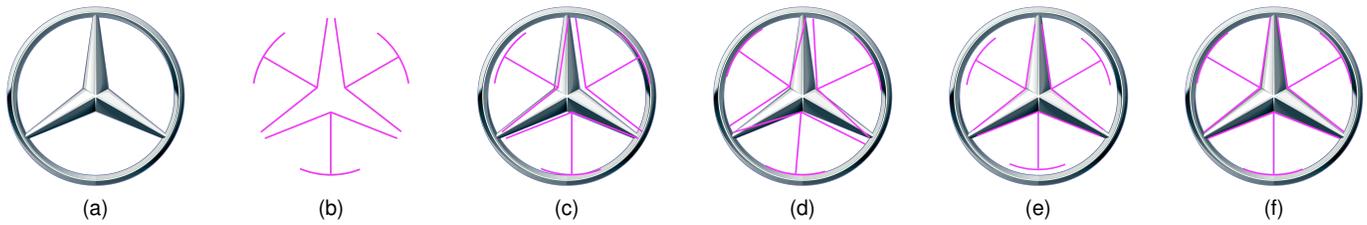

Fig. 2. **Geometric complementary textures** utilise geometrical primitives generated from the shape of a real object. These primitives are not limited to the use of edges, but diagonals, bisectors, inscribed circles, and more complex geometric patterns. In this example, the shape of a real object (a) is used to generate a geometric complementary texture applied to a virtual replica (b). These geometrical primitives allow to identify misalignment errors in position (c), orientation (d), and scale (e) (caused by misalignment in depth). When proper alignment is achieved, this type of texture leads to a seamless observation of the real and virtual objects as if they were part of a single one (f).

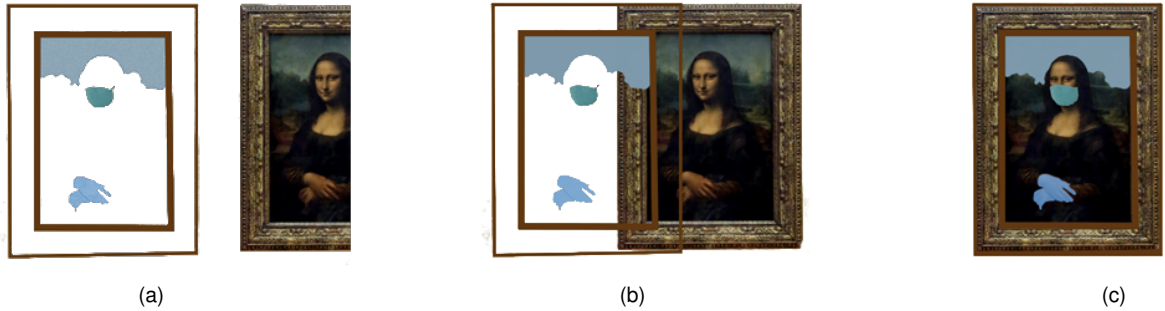

Fig. 3. The **semantic complementary textures** use semantic information derived from the object in the context of the alignment scenario. These semantic complements allow to identify misalignment errors by the visualization of augmented objects that are semantically familiar to the users. In this example, a face mask and a pair of gloves (a) are used to provide semantic information (Covid transmission protection) during the alignment of a frame replica to its targeted painting (b), (c).

During the alignment process, when these components are brought into proximity and misalignment occurs, the result becomes uncomfortable to look at. To this regard, additional edges that arise from the mixing process of the colors create visual noise when using the photometric or geometric complements, while the semantic modality provides context that is broken. This invokes a desire to move the components to remove the noise or semantic absurdity, so that the two parts form a new whole. Thus, we propose new concepts to exploit this natural desire of bringing together these two separate objects into alignment when using the complementary textures in MR applications.

## 4 PROOF OF CONCEPT

To explore the benefits of using COMPLEMENTARY TEXTURES and present a proof-of-concept, we propose an exemplary implementation in which scanned real objects are used to generate a set of complementary textures. These textures, are then applied to virtual replicas of their real counterparts and used during alignment tasks. To illustrate how users could benefit from using these textures during positioning and alignment tasks, we compute visual salience maps of the real and virtual representations. We present them together with respective salience maps of the virtual objects using common visualization techniques frequently observed in MR applications such as wireframe, fresnel and silhouette representations.

### 4.1 Texture Generation

In alignment applications in MR, the user has a virtual model of the object which needs to be aligned with its real counterpart. The virtual models often represent existing 3D models of the object of interest, e.g. designed or scanned CAD models of industrial objects or segmented Computed Tomography or Magnetic Resonance Imaging of organs in medical applications. Depending on the complementary texture modality to be used, the geometrical characteristics, pre-existing textural surface pattern of the object or semantic functionality or properties of the object can be used.

To generate a homogeneous photometric complement, the corresponding pixel colors and brightness values of the preexisting surface pattern of the real object in the form of a UV map can be inverted and applied to the virtual object as shown in Figure (1h). When a non-homogeneous photometric complement is desired, the corresponding alpha values or pixel colors of the original UV map in the region of interest can be modified and applied to the virtual replica as depicted in Figure (1c). The overall procedure to generate a photometric complementary texture can be seen in Figure 5. In practice, it is obvious that one may also need to take the characteristics of AR display into account, in particular considering a color calibration of the display [12].

Alternatively, the geometric complements can be generated by selecting the geometrical structures that provide useful visual cues during the task performance as those depicted in Figure 2. These cues, could also be generated using automatic algorithms that involve mathematical models such as Delaunay triangulation [4], geometric interpolation, fitting curves or even fractals. This type of automatically generated geometrical complements can be particularly interesting when dealing with alignment of architectural elements, as the styles found in several Hindu and Gothic temples [24].

A similar approach can be used for the generation of semantic complementary textures. However, this procedure requires a more selective and less automatic design process as the texture needs to present visual information that is perceptually but also semantically understandable and useful during the alignment task. In Figure 4, a semantic complementary texture is generated from a day scene of the Colosseum. We do not go into details of automatic semantic complementary textures since these would depend a lot on particular scenes and their semantic interpretations, but the concept could be simple and attractive for applications where for example no experts need to provide alignment information.

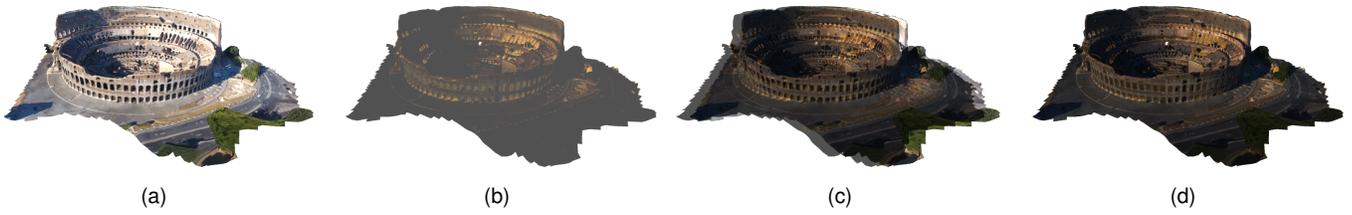

Fig. 4. The use **semantic complementary textures** can be extended to three-dimensional objects. In this example, a model of the Colosseum acquired during the day (a) is used to generate a semantic complement applied to a virtual replica (b) of the building at night. These textures provide contextual information that can be understood by the users during the alignment task (c)-(d). Similarly to other modalities of the complementary textures, the virtual model allows to identify misalignment errors in the form of highly salient visual cues.

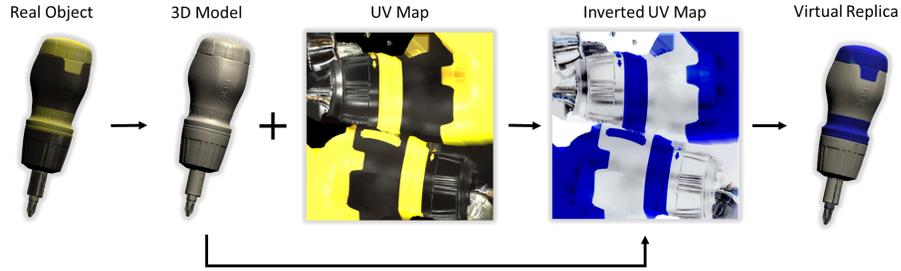

Fig. 5. A **photometric complementary texture** can be generated by inverting the pixel colors and brightness values of the textural surface pattern of a real object in the form of a UV map. This texture can be later applied to the virtual object that will be used during the alignment task.

### 4.2 Alignment Scenario

To explore how the users could benefit from using complementary textures, we implemented an exemplary alignment scenario in a virtual environment using Unity 3D. For the generation and visualization of the objects of interest, we used models of real objects acquired using 3D scanners.[1] The corresponding UV maps generated from the object's scan were later used to generate the specific modality of complementary texture.

To visualize the effects obtained with the different modalities of complementary textures, we used two virtual objects containing the texture of the real object and its respective complementary texture. The alpha value of the virtual replica using the complementary texture was modified to exemplify the use case of MR applications that involve OST-HMDs. Our alignment scenario, allowed to manipulate the position, orientation and scale of the virtual object rendered with the complementary texture, while the real object with its original texture was set to be static.

[1]The model used for Figure 1 and the related figures can be retrieved from https://www.artec3d.com/3d-models.

In addition, we rendered virtual replicas of the textured object using common visualization techniques frequently used for alignment tasks in MR [18]. A representation of these objects can be seen in Figure 6.

### 4.3 Visual Salience

To investigate the visual effects observed when using the complementary textures, as well as traditional visualization techniques, we computed their corresponding salience maps using the Quaternion [25] and Spectral Visual Saliency [26] Toolboxes in Matlab R2018a. An Eigen-based Phase Quaternion Fourier Transform (Eigen-PQFT) approach was selected to compute the salience maps as this type of algorithms outperform the performance of low-level algorithms used for the prediction of human gaze fixation [27].

The resulting salience maps obtained after applying the Eigen-PQFT algorithm to the virtual objects presented in Figure 6 can be seen in Figure 7. From these salience maps, it is possible to see that the use of complementary textures (Figure 7e) provides visual information from both, the exterior and the interior of the object of interest without leading to the observation of a cluttered environment. In contrast, existing approaches mostly provide visual information that is localized on the exterior of the object to align (Figures 7b and 7c), or provide too many details that can end up saturating the visual field (Figure 7d).

To exemplify how users benefit from the complementary textures,

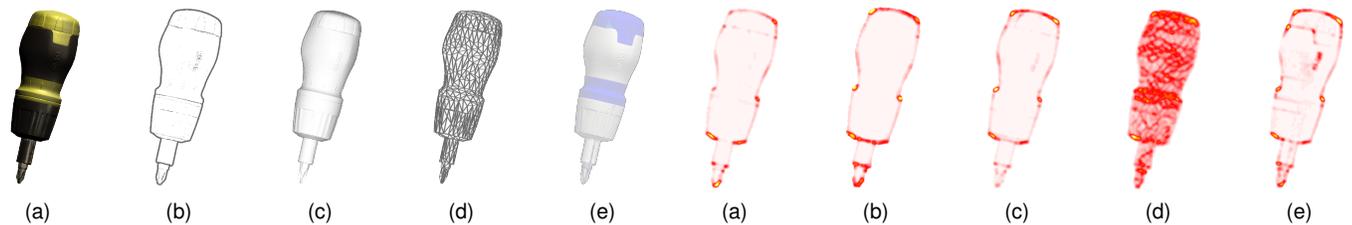

Fig. 6. Virtual objects used during the alignment scenario. The object with the real texture (a) is used as a reference to align virtual replicas using silhouette (b), fresnel (c), wireframe (d), and complementary textures (e).

Fig. 7. Salience maps of the virtual objects used during the alignment scenario. The maps correspond to the real (a), silhouette (b), fresnel (c), wireframe (d), and complementary textures (e).

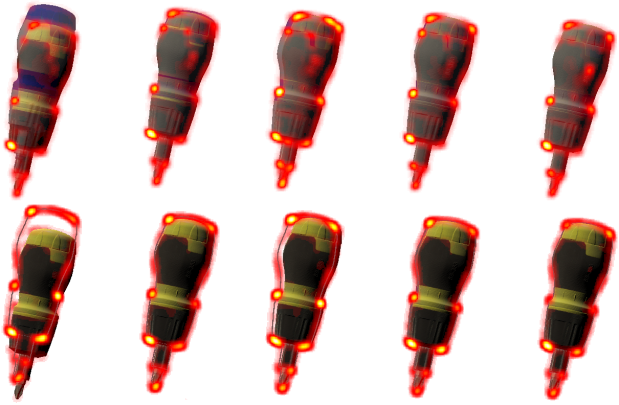

Fig. 8. Exemplary alignment of a screwdriver when errors in translation are observed. Our photometric complementary textures provide variable visual salience values that increases during the alignment process even when small misalignment errors are observed (top). In contrast, the use of classical techniques such as silhouette visualizations provide relatively constant salience values during the alignment (bottom).

we used a similar setup to visualize the salience maps during the alignment process using a homogeneous photometric complementary texture. This texture provides visual areas that increase their salience when the objects are close to be aligned, even when the misalignment errors are small, and decrease once the objects are perfectly aligned. Compared to other visualization techniques such as the silhouette, these highly salient areas remain mostly constant during the alignment process (See Figure 8). Such a behaviour is also depicted in Figure 9, where the integral image and the maximum value of the salience maps computed during the observation of translation and orientation errors are presented. These values, show that the photometric complement presents high integral values in translation and orientation that decrease when perfect alignment is achieved ( Figures 9a and 9b). Moreover, when misalignment between the objects exist, the maximum salience values are observed using this type of visualization (Figures 9c and 9d).

### 4.4 Demonstration

To demonstrate an initial implementation of our approach, we created an exemplary demonstration of the COMPLEMENTARY TEXTURES using Unity 3D and the Microsoft HoloLens OST-HMD. The pair of real and virtual objects selected for this demonstration correspond to an acrylic cube with printed paper textures attached to its surface and edge length of 9.1$cm$. The textures attached to the cube correspond to a simple implementation of photometric complementary textures depicting black and white triangles of different sizes arranged in multiple configurations. The corresponding complementary texture was applied to a virtual cube of the same dimensions as the physical cube. In this scenario, we have chosen to use this object geometry and complementary texture to focus the attention of the observer on the visual cues provided by the object's texture and not by its shape.

As can be seen in Fig. 10, despite the black and white images present a best case scenario for the creation of the complementary texture and result in a low error saliency, matching the real and virtual textures precisely, as what would be necessary if complete loss of hue is desired, presents a very challenging scenario under real conditions. Following this line, the screen used to provide the augmentations would need to be precisely color calibrated [12], and some colors of the real object might not be possible to complement exactly as the color might be out of the display range. Furthermore brightness differences of the real object due to light and shadow would need to be matched.

Nevertheless, and as can also be observed from Fig. 10, even when the contrast, brightness and uniformity of the virtual and real object do not match precisely, the effect of the complementary textures can still work, as the visual salience goes down when the object is in close proximity to its counterpart. This effect is especially observable on the edges between the textures as these regions provide effective visual cues to judge the alignment. Thus, showing that misalignment errors can be perceived by the observer in scenarios under real lighting conditions.

## 5 DISCUSSION

In this paper, we presented the COMPLEMENTARY TEXTURE as a novel concept for assisting users during the alignment of virtual and real objects in MR scenarios. This concept provides users with a meaningful virtual representation that goes beyond the simple presentation of a desired position using classical visualization techniques, but presents additional information that can be used to perceive misalignment errors during task performance.

The proof-of-concept presented in this work, shows that using COMPLEMENTARY TEXTURES provides interactive visual cues that can be perceived by the observer during the full process of the alignment task. The properties of this concept lead to the observation of high salience values even when small misalignment errors exist. Such values decrease when proper alignment is achieved. This property can then be used to improve the alignment of the objects, as the observer is more aware of the errors.

While the use of photometric complementary textures may not be suitable for all possible alignment applications encountered in MR environments, alternative modalities of the COMPLEMENTARY TEXTURES such as the geometric and semantic complements can be used to assist the user. We expect that the use of geometric complements would be beneficial in scenarios that require special attention to objects of interest without leading to visual occlusion, as it is the case of medical applications and industrial settings. Moreover, we believe that the use of semantic complements could be used in alternative scenarios where users do not need to be fully aware of the geometry of the objects but they could achieve proper alignment by reaching a semantically meaningful scene after alignment.

### 5.1 Design Considerations, Applications and Challenges

For the initial proof-of-concept, all complementary textures were designed and implemented using a non-automatic procedure. However, the nature of the photometric complements allows for generating the complements by simply inverting the pixel colors and brightness values of the virtual model of the real object without the necessity of involving computationally complex algorithms. In a similar manner, the geometrical computations, e.g. triangulation, could potentially allow for the partial or completely automatic generation of the COMPLEMENTARY TEXTURES.

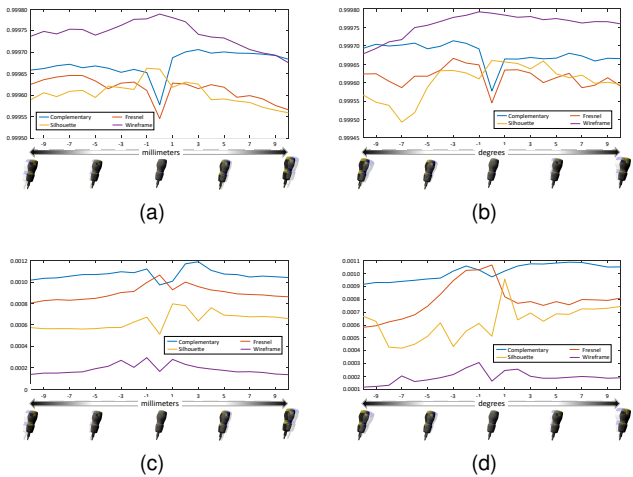

Fig. 9. Integral image (a),(b) and maximum value (c),(d) of the salience maps computed during the observation of translation and orientation errors when using silhouette, fresnel, wireframe and complementary textures.

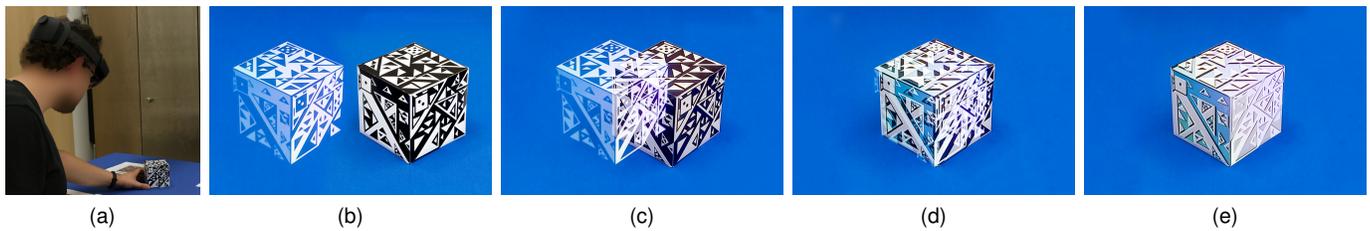

Fig. 10. (b) Real world demonstration of a user aligning a cube with a complementary texture. (b) Depicts the virtual and real object separately, and the process of object alignment ((c) - (d)). The salience observed is perceived as visual noise until the objects are properly aligned (e).

In regards to the geometric complements, like those presented in this work, the geometric single-colored shapes are the easiest to generate. Exemplary automatic procedures could include the extraction of key features based on high curvature textures and/or surfaces. Subsequently, the corresponding complementary texture can be created from its original counterpart. Alternatively, more complex mathematical models based on Delaunay triangulation, geometric interpolation, fitting curves or fractals could be used for the generation of the textures.

In contrast, the creation of automatic semantic complements present a greater challenge as they should provide visual semantic information that can be naturally understood by the user during the alignment task. One can still imagine automatic methods for class of objects, e.g. eyes can be automatically detected for generating matching glasses or words could be detected by automatic recognition systems and be completed such that the alignment of two graphics would semantically make sense. One could even think that future deep learning approaches could support the generation of complementary textures based on the object geometry, texture and even semantic properties.

Here, the design of the complementary textures has been performed assuming the observation of consistent lighting conditions under controlled environments. In this regard, it is important to consider how the different lighting conditions observed in several real environments could influence the way in how the complementary textures interact with their real counterparts. However, we expect that the adaptation of radiometric compensation techniques such as the one presented by Grundhofer and Bimber [8] could contribute to mitigate these effects. The use of geometric complementary textures could represent an alternative for this type of environmental conditions, specially when the conditions of the environment would impede the use of radiometric compensation techniques.

## 6 CONCLUSION

In this paper, we present the COMPLEMENTARY TEXTURES a novel visualization concept for alignment of real and virtual objects in MR applications. The proposed virtual replica design and visualization represent a novel and unexplored alternative to the classical methods that are frequently used in these environments. The different modalities of COMPLEMENTARY TEXTURES presented in this work are manifold and showcase the versatility of this concept. Our proof-of-concept experiments show that the COMPLEMENTARY TEXTURES provide visual cues that generate high salience when real and virtual objects are misaligned. Such salience values decrease when both objects are properly aligned. The promising preliminary results lead us to believe that our concept could open paths for the community to develop suitable and creative alternatives to support users for object alignment in MR, making it easy to use and smooth to integrate into a broad range of solutions.